\documentclass[a4paper,fleqn,usenatbib,useAMS]{mnras}

\usepackage{graphicx}	
\usepackage{amsmath}	
\usepackage{amssymb}	
\usepackage{multicol}        
\usepackage{bm}		
\usepackage{pdflscape}	
\usepackage[version=4]{mhchem}
\usepackage{multirow}
\usepackage{tablefootnote}

\newcommand{\hdo}{{\rm HDO/H_{2}O}}
\newcommand{\ddo}{{\rm D_{2}O/HDO}}
\newcommand{\nhhd}{{\rm NH_{2}D/NH_{3}}}
\newcommand{\nddh}{{\rm ND_{2}H/NH_{2}D}}

\usepackage[T1]{fontenc}
\usepackage{ae,aecompl}

\usepackage{newtxtext,newtxmath}

\title[Tracing the atomic nitrogen abundance with ammonia duteration]{Tracing the atomic nitrogen abundance in star-forming regions with ammonia deuteration}

\author[K. Furuya \& M. V. Persson]{
Kenji Furuya$^{1}$\thanks{E-mail: furuya@ccs.tsukuba.ac.jp}
and Magnus V. Persson$^{2}$
\\
$^{1}$Center for Computational Sciences, University of Tsukuba, 1-1-1 Tennoudai, Tsukuba 305-8577, Japan\\
$^{2}$Department of Space, Earth and Environment, Chalmers University of Technology, Onsala Space Observatory, 439 92, Onsala,
Sweden
}

\date{Last updated; in original form}

\pubyear{2017}

\begin{document}
\label{firstpage}
\pagerange{\pageref{firstpage}--\pageref{lastpage}}
\maketitle

\begin{abstract}
Partitioning of elemental nitrogen in star-forming regions is not well constrained. 
Most nitrogen is expected to be partitioned among atomic nitrogen (\mbox{\ion{N}{i}}), molecular nitrogen (\ce{N2}), 
and icy N-bearing molecules, such as \ce{NH3} and \ce{N2}.
\mbox{\ion{N}{i}} is not directly observable in the cold gas.
In this paper, we propose an indirect way to constrain the amount of \mbox{\ion{N}{i}} in the cold gas of star-forming clouds, 
via deuteration in ammonia ice, the [$\nddh$]/[$\nhhd$] ratio.
Using gas-ice astrochemical simulations, we show that if atomic nitrogen remains as the primary reservoir of nitrogen during cold ice formation stages, 
the [$\nddh$]/[$\nhhd$] ratio is close to the statistical value of 1/3 and lower than unity, 
whereas if atomic nitrogen is largely converted into N-bearing molecules, the ratio should be larger than unity.
Observability of ammonia isotopologues in the inner hot regions around low-mass protostars, where ammonia ice has sublimated, is also discussed.
We conclude that the [$\nddh$]/[$\nhhd$] ratio can be quantified using a combination 
of VLA and ALMA observations with reasonable integration times, at least toward 
IRAS 16293-2422 where high molecular column densities are expected.
\end{abstract}



\section{Introduction}
\label{sec:intro}
One of the most fundamental questions in the field of astrochemistry is how heavy elements are partitioned among chemical species at each evolutionary stage 
during star- and planet-formation.
Nitrogen is the fifth most abundant element with the abundance of [N/H]$_{\rm elem}=6\times10^{-5}$ in the local interstellar medium (ISM) \citep{przybilla08}.
The partitioning of elemental nitrogen in dense star-forming regions is not well constrained.
Determining it is important for several reasons.
It is a key for the understanding of differential behavior of N-bearing species from O- and C-bearing species in the gas phase in prestellar cores and
nitrogen isotope fractionation chemistry \citep[e.g.,][]{bergin07,rodgers08}.
The elemental nitrogen partitioning would also affect the formation efficiency of N-bearing complex organic molecules, as, for example, 
\ce{N2} is more stable than \ce{NH3} \citep[e.g.,][]{daranlot12}.
Furthermore, the nitrogen partitioning in star forming clouds may affect that in protoplanetary discs \citep{schwarz14}, which would shape the composition of planets.

There is no clear evidence of nitrogen depletion into dust grains unlike other elements \citep{jenkins09}.
Then most nitrogen should be present in the gas phase in diffuse clouds.
The dominant form of nitrogen in diffuse clouds is the atomic form (\mbox{\ion{N}{i}}) rather than either the ionic form or molecular nitrogen (N$_2$) \citep{viala86,knath04}.
In dense molecular clouds and cores, nitrogen chemistry consists of three competing processes; 
(i) the conversion of \mbox{\ion{N}{i}} into N$_2$ in the gas phase,
(ii) destruction of N$_2$ via e.g., photodissociation and reaction with He$^+$, and 
(iii) freeze out of \mbox{\ion{N}{i}} and N$_2$ onto dust grains followed by surface reactions \citep[e.g.,][]{herbst73,hidaka11,daranlot12,li13}.
From these, most nitrogen is expected to be partitioned among \mbox{\ion{N}{i}}, N$_2$, 
and icy N-bearing molecules, such as ammonia (NH$_3$) and N$_2$ ices.
The {\it Spitzer} ice survey showed that the ice in star-forming regions contains, on average, $\sim$10 \% of 
overall nitrogen as NH$_3$, NH$_4^+$, and OCN$^{-}$ \citep{oberg11}, 
adopting the water ice abundance of $5\times10^{-5}$ with respect to hydrogen nuclei \citep{boogert04,gibb04,pontoppidan04}.
The remaining nitrogen would be present in the gas phase mainly as \mbox{\ion{N}{i}} or N$_2$.
Alternatively, there is still the possibility that a large fraction of nitrogen is locked up in N$_2$ ice.
Indirect measurements of the upper limits on the amount of N$_2$ ice in molecular clouds are rather uncertain, 
$\lesssim$40 \% of overall nitrogen \citep{elsila97,boogert02,boogert15}.

Gas-phase nitrogen chemistry that converts \mbox{\ion{N}{i}} into N$_2$ is different from oxygen and carbon chemistry, 
in which formation of molecules from atoms are initiated by reactions with \ce{H3+}.
It has been thought that N$_2$ forms via neutral-neutral reactions as follows \citep[e.g.,][]{herbst73,hilyblant10,daranlot12}:
\begin{align}
\ce{N}+\ce{CH} &\rightarrow \ce{CN}, \label{react:n+ch}\\
\ce{CN} + \ce{N} &\rightarrow \ce{N2}, \label{react:n+cn}
\end{align}
and
\begin{align}
\ce{N} + \ce{OH} &\rightarrow \ce{NO},\\
\ce{NO} + \ce{N} &\rightarrow \ce{N2}. \label{react:no+n}
\end{align}
According to gas-phase astrochemical models for dense clouds and cores, where the interstellar UV radiation field is significantly attenuated,
either \ce{N2} or \mbox{\ion{N}{i}} can be the biggest nitrogen reservoir in steady-state, 
depending on the C/O elemental ratio and the elemental abundance of sulfur that are available for gas phase chemistry \citep{hilyblant10,legal14}.
Both the C/O ratio and the sulfur abundance would be time-dependent due to ice formation in reality.
Recent gas-ice astrochemical simulations \citep{furuya15,furuya16}, which trace the physical and chemical evolution from 
translucent clouds to denser cores, have predicted that 
N$_2$ is the primary form of nitrogen in the gas phase in cold dense cores, 
while most of elemental nitrogen exists as ice in the forms of \ce{NH3} and \ce{N2} \citep[see also][]{maret06,daranlot12}.

Neither \mbox{\ion{N}{i}} nor N$_2$ in the gas phase is directly observable in dense cores due to the cold temperatures.
Several observational studies have attempted to constrain the abundance of N$_2$ in the gas phase via a proxy molecule, N$_2$H$^+$ which is primary formed by N$_2$ + H$_3^+$.
\citet{maret06} inferred the gaseous N$_2$ abundance in dense cloud B68 from the observations of N$_2$H$^+$ supplemented by CO and HCO$^+$ observations, 
which were used to constrain the abundances of main destroyers for \ce{N2H+}, i.e., CO and electrons.
They concluded that the gaseous N$_2$ abundance is low (a few \%) compared to overall nitrogen abundance of $6\times10^{-5}$, 
based on chemical and radiative transfer models \citep[see also][]{pagani12}.
The result is consistent with earlier investigation of the N$_2$ abundance in the gas of several cold dense clouds estimated from N$_2$H$^+$ observations with simpler analysis.
\citet{mcgonagle90} estimated the gaseous \ce{N2}/\ce{CO} ratio of $\sim$8 \% in dark cloud L134N, assuming that 
the ratio is close to the \ce{N2H+}/\ce{HCO+} ratio, i.e., destruction of \ce{N2H+} by CO is neglected.
If the canonical CO abundance of 10$^{-4}$ is adopted, the gaseous \ce{N2} abundance is evaluated to be $\sim 8\times10^{-6}$ with respect to hydrogen nuclei, 
corresponding to $\sim$30 \% of overall nitrogen.
\citet{womack92} estimated the gaseous \ce{N2} abundance of four cold clouds to be $\sim$3$\times10^{-6}$ with respect to hydrogen nuclei, 
corresponding to $\sim$10 \% of overall nitrogen.

\citet{hilyblant10} suggested that the \mbox{\ion{N}{i}} abundance in the gas phase can be constrained from the CN/HCN abundance ratio. 
The basic idea (or assumption) behind this is that both the production rate and destruction rate of CN depend on the \mbox{\ion{N}{i}} abundance (see Reactions (1) and (2)), 
while for HCN, only the production rate depends on the \mbox{\ion{N}{i}} abundance.
Based on the observationally derived CN/HCN ratio in several prestellar cores, it was suggested that the \mbox{\ion{N}{i}} abundance is low, 
up to a few \% of overall nitrogen.
The method, however, suffers from uncertainties of formation and destruction pathways of relevant species and their rate coefficients \citep{hilyblant10}.
Indeed, their astrochemical models failed to explain the observationally derived abundances of N-bearing molecules in the prestellar cores.

Thus previous studies of the nitrogen budget exploration are inconclusive, while it seems unlikely that most elemental nitrogen exists as gaseous \ce{N2}.
In this paper, we propose a new way to indirectly trace the evolution of the \mbox{\ion{N}{i}} abundance in the cold gas of star-forming regions via deuteration of ammonia ice.
The paper is organized as follows.
A proposed method is presented through a simple analytical model in Section \ref{sec:theory}, 
while in Section \ref{sec:model}, the method is verified by gas-ice astrochemical simulations.
Deuteration measurements of the ISM ice relies heavily on the gas observations toward inner warm ($>$100 K) regions 
in the deeply embedded protostars, 
where ices have sublimated.
The observability of deuterated ammonia in the warm gas is discussed in Section \ref{sec:discussion}.
Our findings are summarized in Section \ref{sec:summary}.

\section{Multiple deuteration of ammonia ice}
\label{sec:theory}
Infrared ice observations have shown that ice formation (at least water ice formation) already starts before the dense core stage, 
where the cores gravitationally collapse to form protostars \citep[e.g.,][]{whittet93,murakawa00}.
Ice formation in star-forming regions can be roughly divided into two stages 
(or  ice mantles have two layered structure in terms of their molecular compositions) \citep[e.g.,][]{pontoppidan06,oberg11}; 
in the early stage, \ce{H2O}-dominated ice layers are formed.
In the later stage, at higher extinction and density, the catastrophic CO freeze out happens, and 
ice layers which mainly consist of CO and its hydrogenated species (\ce{H2CO} and \ce{CH3OH}) are formed.
Observations toward deeply embedded low mass protostars have revealed that the level of methanol deuteration, in particular the \ce{CH3OD}/\ce{CH3OH} ratio, 
is much higher than the HDO/\ce{H2O} ratio in the warm ($\gtrsim$100 K) gas around the protostars, where ices have sublimated 
\citep[$\sim$10$^{-2}$ versus $\sim$10$^{-3}$,][]{parise06,persson13,persson14,coutens14}.
This trend indicates that deuterium fractionation is more efficient in the later stage of the ice formation when the catastrophic CO freeze out happens  \citep{cazaux11,taquet12,taquet14,furuya16}. 

It is thought that ammonia ices are primary formed via sequential hydrogenation/deuteraion of atomic nitrogen on a (icy) grain surface,
supported by laboratory experiments \citep{hidaka11,fedoseev15a,fedoseev15b}.
We show how the evolution of atomic nitrogen abundance during ice formation is reflected in the [$\nddh$]/[$\nhhd$] ratio, 
using a simple analytical model.
Specifically, one can distinguish two cases, whether the significant fraction of elemental nitrogen is present in the atomic form until the later stage of ice formation 
or not, using the [$\nddh$]/[$\nhhd$] ratio.
A similar analysis was made by \citet{furuya16} for water ice to explain the higher \ce{D2O}/HDO ratio than the HDO/\ce{H2O} ratio 
observed in the warm gas around a protostar \citep[$\sim$10$^{-2}$ versus $\sim$10$^{-3}$;][]{coutens14}.

Let us consider a two stage model (or a two-layer ice model).
We denote the total amount of nitrogen locked into \ce{NH3}, \ce{NH2D}, or \ce{ND2H} ices at each stage (or in each layer) $k$, where $k$ = I or II, as $N_k$.
Denoting the fraction of nitrogen locked into \ce{NX3} ice, where X is H or D, as $P_{\ce{NX3},\,\,k}$,
one can express the amount of \ce{NX3} ice formed in stage $k$ as $P_{\ce{NX3},\,\,k}N_{k}$.
We introduce a free parameter $q$ that satisfies $P_{\ce{ND2H},\,\,k}$/$P_{\ce{NH2D},\,\,k}$ = $qP_{\ce{NH2D},\,\,k}$/$P_{\ce{NH3},\,\,k}$.
\citet{fedoseev15b} found that sequential reactions of H and D atoms with atomic nitrogen on a cold substrate lead to 
the [$\nddh$]/[$\nhhd$] production rate ratio of $\sim$1/3, i.e., the statistical ratio, 
in their experiments.
On the other hand, they found that the production rates for \ce{NH3}:\ce{NH2D}:\ce{ND2H}:\ce{ND3} (0.4:2.1:3.5:2) are deviated from the statistical distribution of 1:3:3:1, 
assuming the atomic D/H ratio of unity in the mixed atom beam fluxes. 
This result indicates that every deuteration reaction has a probability of a factor of $\sim$1.7 higher to occur over the corresponding hydrogenation reactions \citep{fedoseev15b}.
They concluded that the main reason for this deuterium enrichment is higher sticking probability of D atoms than that of H atoms in their experiments 
(i.e., the atomic D/H ratio on a surface is higher than unity).
Indeed, the production rates obtained by their experiments can be explained by the statistical distribution with a H:D ratio of 1:1.7, 
i.e., \ce{NH3}:\ce{NH2D}:\ce{ND2H}:\ce{ND3} = 0.4:2.1:3.5:2 = 1:5.3:8.8:5 $\sim$ 1:$1.7 \times 3$:$(1.7)^2 \times 3$:$(1.7)^3 \times 1$.
It should be noted that the statistical distribution of \ce{NH3}:\ce{NH2D}:\ce{ND2H}:\ce{ND3} does depend on the atomic D/H ratio, 
while the statistical ratio of [$\nddh$]/[$\nhhd$] does not.
This characteristic makes the [$\nddh$]/[$\nhhd$] ratio more useful for investigating nitrogen chemistry than the \ce{NH3}:\ce{NH2D}:\ce{ND2H} distribution \citep[cf.][]{rodgers02}.
We assume $q=1/3$.

Using the above relations and denoting the $\nhhd$ ratio and the $\nddh$ ratio as $f_{\rm D1}$ and $f_{\rm D2}$, respectively, 
one can express the $f_{\rm D2}$/$f_{\rm D1}$ ratio in the whole ice mantle as follows \citep[cf. Eq. 4 in][]{furuya16}:
\begin{align}
\frac{f_{\rm D2}}{f_{\rm D1}} &= \frac{\sum_{k={\rm I,\,II}}P_{\ce{ND2H},\,\,k}N_{k} \bigg/ \sum_{k={\rm I,\,II}}P_{\ce{NH2D},\,\,k}N_{k}}
{\sum_{k={\rm I,\,II}}P_{\ce{NH2D},\,\,k}N_{k} \bigg/ \sum_{k={\rm I,\,II}}P_{\ce{NH3},\,\,k}N_{k}},\\
&\approx \frac{(1+\alpha)(1+\alpha \beta^2)}{3(1+\alpha \beta)^2},
\end{align}
where parameter $\beta$ is defined as $P_{\rm {\ce{NH2D},\,\,II}}/P_{\rm {\ce{NH2D},\,\,I}}$, 
and $P_{\ce{NH3},\,\,k}$ is assumed to be $\sim$1.
Parameter $\alpha$ is defined as $N_{\rm N,\,\, II}/N_{\rm N,\,\, I}$ (i.e., amount of gaseous \mbox{\ion{N}{i}} in the later stages versus early stages).
Then $0 < \alpha < 1$ corresponds to the case when most \ce{NH3} ice is formed in the early stage,
but the production of ammonia ice continues in the later stage with reduced efficiency.
This can be interpreted that most \mbox{\ion{N}{i}} is consumed by the formation of N-bearing molecules, such as \ce{N2} and \ce{NH3}, in the early stage, 
and only a small amount of \mbox{\ion{N}{i}} remains in the later stage.
$\alpha \geq 1$ corresponds to the case when most \ce{NH3} ice is formed in the later stage,
i.e.,  gaseous \mbox{\ion{N}{i}} (the source of \ce{NH3} ice) remains as the primary nitrogen reservoir until the later stage, 
and the production of \ce{NH3} ice is enhanced due to increased density and attenuation of UV radiation.

\begin{figure}
\resizebox{\hsize}{!}{\includegraphics{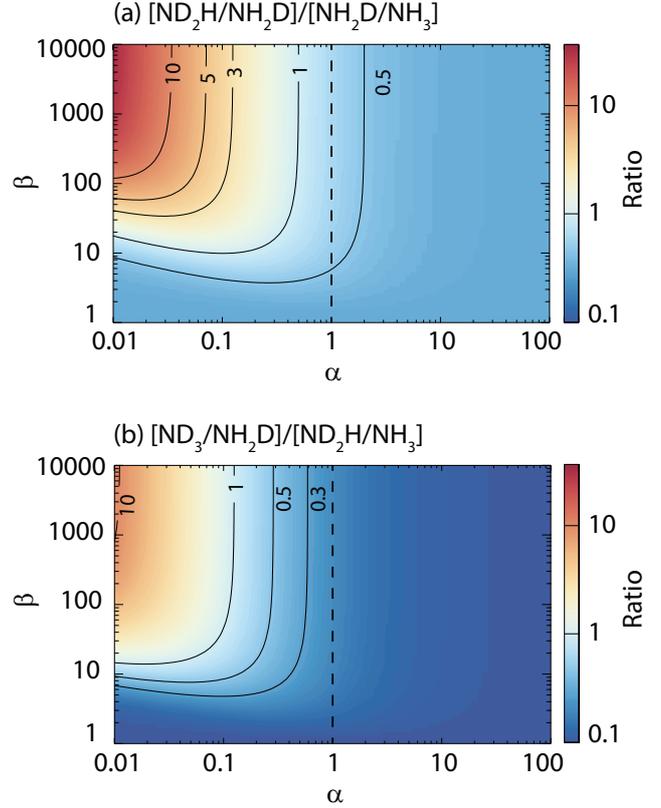}}
\caption{The [\ce{ND2H}/\ce{NH2D}]/[\ce{NH2D}/\ce{NH3}] ratio (top panel) and the [\ce{ND3}/\ce{NH2D}]/[\ce{ND2H}/\ce{NH3}] ratio (bottom panel) 
as functions of  $\alpha$ (= $N_{\rm II}/N_{\rm I}$) and $\beta$ (= $P_{\rm {\ce{NH2D},\,\,II}}/P_{\rm {\ce{NH2D},\,\,I}}$) in the analytical two stage model.
The vertical dashed lines indicate $\alpha$ of unity.
See the text for more information.
}
\label{fig:analytic}
\end{figure}

\begin{figure}
\includegraphics[width=\columnwidth]{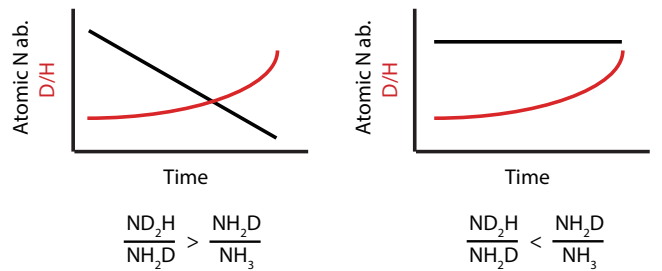}
\caption{
Left) In the case when most of atomic nitrogen is converted into molecules during ice formation, the [\ce{ND2H}/\ce{NH2D}]/[\ce{NH2D}/\ce{NH3}] ratio in the whole ice mantle is
greater than unity. Right) In the case when atomic nitrogen is the primary nitrogen reservoir during the ice formation stages, the [\ce{ND2H}/\ce{NH2D}]/[\ce{NH2D}/\ce{NH3}] ratio is similar to the statistical ratio of 1/3 and lower than unity.
}
\label{fig:cartoon}
\end{figure}

Figure \ref{fig:analytic}, the top panel shows the $f_{\rm D2}$/$f_{\rm D1}$ ratio as a function of $\alpha$ and $\beta$.
It is sufficient to consider the case of $\beta > 1$, because the deuterium fractionation becomes more efficient in the later stage.
Note that to explain $\sim$10 times higher \ce{D2O}/HDO ratio than the HDO/\ce{H2O} ratio, $\beta \gtrsim 100$ is required for HDO \citep{furuya16}.
For $\alpha \lesssim 0.5$ and $\beta \gtrsim 10$, the $f_{\rm D2}/f_{\rm D1}$ ratio is higher than unity and increases with decreasing $\alpha$.
For $\alpha \gtrsim 0.5$, the $f_{\rm D2}/f_{\rm D1}$ ratio is lower than unity, regardless of $\beta$.
Then the ratio allows us to prove the evolution of the \mbox{\ion{N}{i}} abundance.
The situation is summarized in Figure \ref{fig:cartoon}.

The above discussion may be an oversimplification.
For example, the gas-phase formation of ammonia followed by freeze-out was neglected.
In the rest of this paper the robustness of the method is tested using astrochemical simulations.

In order to find the best tracer of \mbox{\ion{N}{i}}, we explored other combinations of ammonia deuteration ratios than the $f_{\rm D2}$/$f_{\rm D1}$ ratio,
using the two-stage model.
Our conclusion is that the $f_{\rm D2}$/$f_{\rm D1}$ ratio is the best.
For example, the [\ce{ND3}/\ce{NH2D}]/[\ce{ND2H}/\ce{NH3}] ratio shows the similar dependence on $\alpha$ and $\beta$ (Figure \ref{fig:analytic}, bottom panel), 
assuming the statistical ratio, $P_{\ce{ND3},\,\,k}$/$P_{\ce{ND2H},\,\,k}$ = $P_{\ce{ND2H},\,\,k}$/$3P_{\ce{NH2D},\,\,k}$.
It is, however, required to additionally detect triply deuterated ammonia, \ce{ND3}.
Then we focus on the $f_{\rm D2}$/$f_{\rm D1}$ ratio in this paper.

\section{Numerical setup}
\label{sec:model}
We simulate molecular evolution from the formation of a molecular cloud to a protostellar core as in our previous studies \citep{furuya15,furuya16}.

\subsection{Physical model}
We use two types of physical models: 
one dimensional shock model for the formation and evolution of a molecular cloud 
due to the compression of diffuse \mbox{\ion{H}{i}} gas by super-sonic accretion flows \citep{bergin04,hassel10} 
and one-dimensional radiation hydrodynamics simulations for the evolution of a prestellar core to form a protostar via the gravitational collapse \citep{masunaga00}.  
Here we present brief descriptions of the two models, while more details can be found in the original papers.

The shock model simulates the physical evolution of post-shock materials, i.e., molecular cloud, 
considering heating and cooling processes in a plane-parallel configuration.
The interstellar UV radiation field of \citet{draine78} is adopted. 
The cosmic-ray ionization rate of H$_2$ is set to be $1.3\times10^{-17}$ s$^{-1}$.
We assume the pre-shock \mbox{\ion{H}{i}} gas density of 10 cm$^{-3}$ and the pre-shock velocity of 15 km s$^{-1}$ as in \citet{furuya15}.
The forming and evolving cloud has the gas density of $\sim$10$^4$ cm$^{-3}$ and gas and dust temperatures 
of $\sim$10-20 K, depending on time (Figure \ref{fig:layer}, panel a).
The column density of the cloud increases linearly with time and the time it takes for the column density to reach $A_V$ = 1 mag is $\sim$4 Myr.
In the shock model, ram pressure dominates the physical evolution rather than self-gravity.

The collapse model simulates the gravitational collapse of a prestellar core with the mass of 3.9 $M_{\odot}$ assuming spherical symmetry.  
The protostar is born at the core center at $2.5 \times 10^5$ yr after the beginning of the collapse, 
corresponding to 1.4$t_{\rm ff}$, where $t_{\rm ff}$ is the free-fall time of the initial central density of hydrogen nuclei $\sim$6$\times10^4$ cm$^{-3}$.
After the birth of the protostar, the model further follows the physical evolution for $9.3 \times 10^4$ yr.

\subsection{Chemical model}
In the physical models, Lagrangian fluid parcels are traced. 
The kinetic rate equation is solved along the fluid parcels to obtain the molecular evolution in the fluid parcels \citep[e.g.,][]{aikawa13}.
The molecular abundances obtained by the cloud formation model are used as the initial abundances for the gravitational collapse model.

Our chemical model is basically the same as that used in \citet{furuya15}.
The gas-ice chemistry is described by a three-phase model, 
in which three distinct phases, gas-phase, icy grain surface, and the bulk of ice mantle are considered \citep{hasegawa93}.
Gas-phase reactions, gas-surface interactions, and surface reactions are considered, while the bulk ice mantle is assumed to be chemically inert.
We consider the top four monolayers of the ice as a surface following \citet{vasyunin13}.
Our chemical network is originally based on that of \citet{garrod06}, 
while gaseous nitrogen chemistry has been updated following recent references \citep[][]{wakelam13,loison14,roueff15}.
The network has been extended to include up to triply deuterated species and nuclear spin states of \ce{H2}, \ce{H3+}, and their isotopologues.
The state-to-state reaction rate coefficients for the \ce{H2}-\ce{H3+} system are taken from \citet{hugo09}.
The self-shielding of \ce{H2}, HD, CO, and \ce{N2} against photodissociation are taken into account \citep[e.g.,][]{visser09,li13}.
The photodesorption yield per incident FUV photon is $\sim3\times10^{-4}$ for water ice \citep{arasa15}.
We assume the yield of $10^{-3}$ for ammonia ice.
The photodesorption yields for CO and \ce{N2} ices are treated in a special way and given as 
increasing functions of the surface coverage of CO \citep{furuya15}.
The yield for CO ice varies from $3\times10^{-4}$ to 10$^{-2}$, while that for \ce{N2} ice varies from $3\times10^{-3}$ to $8\times10^{-3}$
 \citep{fayolle11,bertin12,bertin13}.

There are some updates from the model in \citet{furuya15};
(i) the treatment of charge exchange and proton/deuteron transfer reactions,
(ii) the binding energy on a grain surface, and
(ii) mass for calculating transmission probabilities of tunneling reactions on grain surfaces.
In our previous work, complete scrambling of protons and deuteron was assumed for all types of reactions \citep{aikawa12}.
The assumption on the complete scrambling is not appropriate, at least, for charge exchange and proton/deuteron transfer reactions \citep{rodgers96,sipila13}.
We relax the assumption for these two types of reactions, following \citet{sipila13}.
This modification reduces the production rates of multiply deuterated species, such as \ce{ND2H} and \ce{ND3}, via gas-phase reactions 
when deuterium fractionation proceeds significantly and multiply deuterated \ce{H3+} becomes abundant.

Laboratory experiments have shown that the binding energy of species depends on the type of surfaces \citep[e.g.,][]{fayolle16}.
As described in Section \ref{sec:theory}, the surface composition of the ISM ice can vary from a \ce{H2O}-rich polar surface to a \ce{CO}-rich apolar surface, depending on time.
In our model the binding energy of species $i$, $E_{\rm des}(i)$, is calculated as a function of surface coverage of species $j$, $\theta_j$, where $j$ = \ce{H2}, CO, \ce{CO2}, or \ce{CH3OH}:
\begin{equation}
E_{\rm des}(i) = (1-\Sigma_j\theta_j)E^{\ce{H2O}}_{\rm des}(i) + \Sigma_j\theta_j E^{j}_{\rm des}(i), \label{eq:edes}
\end{equation}
where $E^j_{\rm des}(i)$ is the binding energy of species $i$ on species $j$.
The set of the binding energies on a water ice substrate, $E^{\ce{H2O}}_{\rm des}$, is taken from \citet{collings04,garrod06,wakelam17}.
In particular for this work,  $E^{\ce{H2O}}_{\rm des}$ for \mbox{\ion{O}{i}}, \mbox{\ion{N}{i}}, CO, and \ce{N2} are set to be 1320 K, 720 K, 1300 K, and 1170 K, respectively, 
following laboratory experiments \citep{fayolle16,minissale16}.
 $E^{\ce{H2O}}_{\rm des}$ for atomic hydrogen and \ce{H2} are set to be 550 K.
There is no laboratory data or estimate for most $E^{j}_{\rm des}(i)$ in the literature.
In order to deduce $E^{j}_{\rm des}$ for all species, where $j$ is either \ce{H2}, CO, \ce{CO2}. or \ce{CH3OH}, we assume scaling relations \citep[cf.][]{taquet14},
\begin{align}
E^{j}_{\rm des}(i) = \epsilon_j E^{\ce{H2O}}_{\rm des}(i),
\end{align}
where $\epsilon_j$ is $E^{j}_{\rm des}(j)/E^{\ce{H2O}}_{\rm des}(j)$.
We adopt $\epsilon_{\ce{H2}}$ = 23/550, $\epsilon_{\ce{CO}}$ = 855/1300, $\epsilon_{\ce{CO2}}$ = 2300/2690, and $\epsilon_{\ce{CH3OH}}$ = 4200/5500 \citep[e.g.,][]{oberg05,cuppen07,noble12}.
\citet{fayolle16} found that the binding energy ratio of  \ce{N2} to CO is around 0.9 regardless of a type of a substrate in their laboratory experiments.
This partly supports the above assumption that $\epsilon_j$ is independent of adsorbed species $i$.
The energy barrier against surface diffusion by thermal hopping is set to be 0.45$E_{\rm des}$ for all species.

The sticking probability of neutral species $i$ on a grain surface is given as a function of the binding energy and dust temperature:
\begin{align}
S(i) = (1-\Sigma_j\theta_j)S'(T_{\rm dust}, E^{\ce{H2O}}_{\rm des}(i)) + \Sigma_j\theta_jS'(T_{\rm dust}, E^{j}_{\rm des}(i)), \label{eq:stick}
\end{align}
where $S'$ is the sticking probability formula recommended by \citet{he16} (see their Eq. 1),
who experimentally investigated the sticking probability for stable molecules on nonporus amorphous water ice.
Equation (\ref{eq:stick}) gives, in general, the sticking probability of around unity for all species at $\sim$10 K, 
while the probability is close to zero well above the sublimation temperature of each species.

Grain-surface reactions are assumed to occur by the Langmuir-Hinshelwood mechanism.
When the rates of surface reactions with activation energy barriers are calculated, reaction-diffusion competition is considered \citep[e.g.,][]{chang07,garrod11}.
The activation energy barriers are assumed to be overcome either thermally or via quantum tunneling, whichever is faster.
The transmission probability of tunneling reactions is calculated assuming a rectangular potential barrier, using the reduced mass of the system \citep{tielens82}.
As pointed out by several authors, the use of the reduced mass of the system for hydrogen abstraction reactions, such as OH + \ce{H2} $\rightarrow$ \ce{H2O} + H, 
underestimates the transmission probability by orders of magnitude \citep{hidaka09,oba12,lamberts14}.
We use the mass of a hydrogen (deuterium) atom instead of the reduced mass in the calculation of the transmission probability of hydrogen (deuterium) abstraction reactions.

\subsection{Parameters}
Elemental abundances adopted in this work are the so called low metal abundances and taken from \citet{aikawa99}.
The elemental abundances relative to H are $7.9\times10^{-5}$ for C, $2.5\times10^{-5}$ for N, and $1.8\times10^{-4}$ for O.
Deuterium abundance is set to be $1.5\times10^{-5}$ \citep{linsky03}.
Our model also includes He, S, Si, Fe, Na, Mg, P, and Cl.
All the elements, including H and D, are initially assumed to be in the form of either neutral atoms or atomic ions, depending on their ionization potential.
Then the ortho-to-para ratio of \ce{H2} is calculated explicitly in our model without making an arbitrary assumption on its initial value.
For initial molecular abundances of the collapse model, we use the molecular abundances at the epoch when the column density reaches 2 mag 
in the cloud formation model.

Before the onset of the core collapse, we assume that the prestellar core keeps its hydrostatic structure for 10$^6$ yr ($\sim$5.6$t_{\rm ff}$).
The visual extinction at the outer edge of the core is set to be 5 mag, being irradiated by the interstellar radiation field of \citet{draine78}.
The cosmic-ray ionization rate of H$_2$ is set to be $1.3\times10^{-17}$ s$^{-1}$ throughout the simulations, 
while the flux of FUV photons induced by cosmic-rays is set to be $3\times10^3$ cm$^{-2}$ s$^{-1}$.

\section{Simulation results}
\subsection{Ice chemistry}
\begin{figure*}
\resizebox{\hsize}{!}{\includegraphics[angle=90]{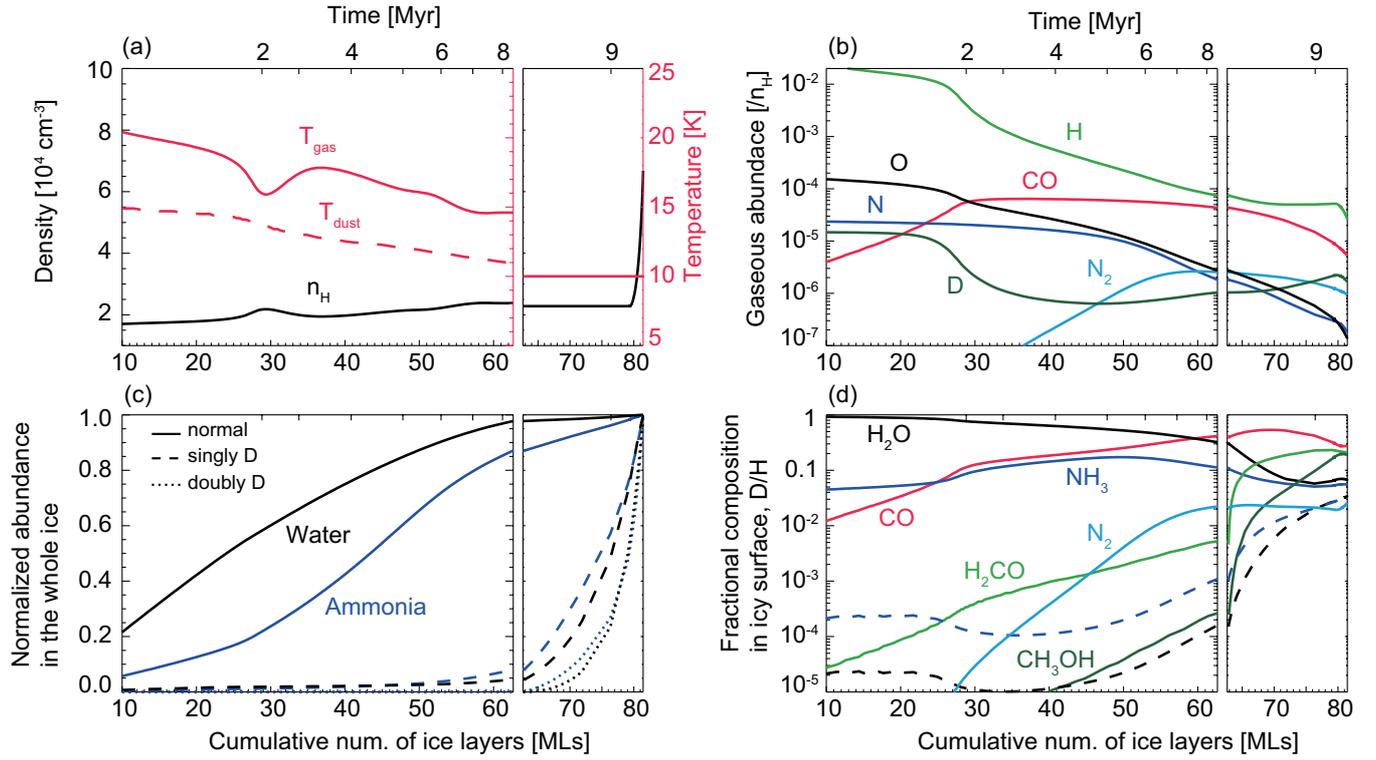}}
\caption{
(a) Physical parameters , (b) abundances of selected gaseous species with respect to hydrogen nuclei, and 
(d) fractional composition in the surface ice layers as functions of the cumulative number of ice layers.
The dashed lines in panel (d) indicate the $\hdo$ ratio (black) and the $\nhhd$ ratio (blue).
As the total number of ice layers increase with time during the cold ice formation stages, the horizontal axis can be read as time (see the label at the top).
In each panel, $\lesssim$60 monolayers (MLs) corresponds to the formation stage of a molecular cloud.
$\gtrsim$60 MLs corresponds to the 1 Myr static core phase and the early collapse phase ($2.6\times10^5$ yr after the onset of collapse), where dust temperature is 10 K.
The results are shown for the fluid parcel, which reaches 5 AU from the central star at the final time of the simulation.
Panel (c) shows the abundances of non-deuterated and deuterated forms of water ice (black) and ammonia ice (blue) 
normalized by their maximum abundances.
Note that the visual extinction for calculating photochemical rates increases from 2 mag at the end of the cloud formation stage to $>$5 mag at the core stage.
}
\label{fig:layer}
\end{figure*}

Figure \ref{fig:layer} shows abundances of selected gaseous species with respect to hydrogen nuclei (panel b) and 
the fractional composition in the surface ice layers (panel d) as functions of the cumulative number of layers formed in the ice mantle.
In each panel, $\lesssim$60 monolayers (MLs) correspond to the cloud formation stage, 
while $\gtrsim$60 MLs correspond to the core stage.
For the core stage, the results for the fluid parcel which reaches 5 AU from the central star at the final simulation time are shown.
The label at the top represents time, in which $t=0$ corresponds to the time when the fluid parcel passes through the shock front, i.e., the onset of cloud formation.

\ce{H2O} ice is the dominant component of the lower ice mantles ($\lesssim$60 MLs) formed in the early times,
while the upper ice layers ($\gtrsim$60 MLs) mainly consists of CO and its hydrogenated species, \ce{H2CO} and \ce{CH3OH}.
The significant freeze-out of CO happens later than the formation of most of water ice in our model, because
the freeze-out of CO is a self-limited process \citep{furuya15};
with increasing the CO coverage on a surface, the photodesorption yield of CO is enhanced and the binding energy of CO is reduced, the latter of which enhances 
the non-thermal desorption rate due to the stochastic heating by cosmic rays.
The conversion of CO ice to \ce{CH3OH} ice becomes more efficient with increasing the abundance ratio of H atoms to CO in the gas phase \citep[$\gtrsim$70 MLs; e.g.,][]{charnley97}.
The $\hdo$ ratio and the $\nhhd$ ratio are much higher in the upper ice mantles than those in the lower ice mantles, due to the CO freeze-out, 
the drop of the ortho-to-para ratio of \ce{H2}, and the attenuation of UV radiation field \citep{furuya16}.

\subsection{Elemental nitrogen partitioning}
\label{sec:n_partition}
At the end of ice formation, when the number of ice layers reaches $\sim$80 MLs,
most elemental nitrogen is distributed among \ce{NH3} ice (76 \%), gaseous and icy \ce{N2} (22 \%), and \mbox{\ion{N}{i}} ($<$1 \%).
In our model, \ce{NH3} ice forms via sequential surface reactions, 
\begin{align}
&\ce{N} + \ce{H} \rightarrow \ce{NH}, \\
&\ce{NH} + \ce{H} \rightarrow \ce{NH2}, \\
&\ce{NH2} + \ce{H2} \rightarrow \ce{NH3} + \ce{H}. \label{react:nh+h2}
\end{align}
The last reaction has an activation energy barrier of 2700 K in the gas phase \citep{Espinosa-Garcia10,hidaka11},
and the transmission probability of this barrier-mediate reaction is set to be $\sim$10$^{-9}$.
Note that reaction-diffusion competition is considered in our model.
In our surface chemical network, the barrierless reaction \ce{NH2} + \ce{H} $\rightarrow$ \ce{NH3} is included, in addition to Reaction (\ref{react:nh+h2}).
Neverthless, Reaction (\ref{react:nh+h2}) is more efficient in our model,
because the adsorption rate of \ce{H2} on grain surfaces is much higher than that of atomic H, and thus \ce{H2} is much more abundant than atomic H on surfaces.
The rate-limiting step of the \ce{NH3} ice formation is the adsorption of \mbox{\ion{N}{i}} on a surface, 
and thus the formation rate of \ce{NH3} ice should not depend on the surface reaction rates significantly.
The \ce{NH3} ice formation is balanced with the photodesorption of \ce{NH3} ice in the cloud formation stage,
and then the timescale of \ce{NH3} ice formation is much longer than the freeze out timescale of \mbox{\ion{N}{i}} ($\sim$10$^5$ yr for the density of 10$^4$ cm$^{-3}$).
In the core stage, the interstellar UV radiation is attenuated significantly, and the timescale of \ce{NH3} ice formation is determined by the freeze out of \mbox{\ion{N}{i}}.

\begin{figure}
\resizebox{\hsize}{!}{\includegraphics{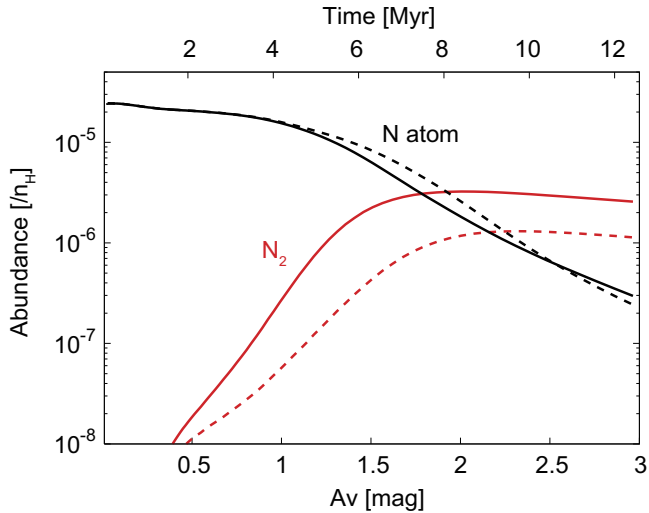}}
\caption{Abundances of atomic nitrogen and \ce{N2} in the cloud formation model.
For \ce{N2}, the total abundance (gas and ice are combined) is shown.
Solid lines represent our fiducial model in which the product of \ce{NH3} photodesorption is set to be \ce{NH3}, 
while dashed lines represent the model in which the products of \ce{NH3} photodesorption are set to be N + 3H.
For this figure, the models of the cloud formation were run until the column density of the cloud reaches 3 mag.
}
\label{fig:nh3_phd}
\end{figure}

Interestingly, the formation of \ce{NH3} ice helps the conversion of \mbox{\ion{N}{i}} into \ce{N2} in the gas phase through photodesorption;
photodesorbed \ce{NH3} is further photodissociated in the gas phase and the photofragment \citep[\ce{NH2} or \ce{NH};][]{heays17} reacts 
with \mbox{\ion{N}{i}} to form \ce{N2}.
\ce{NH} and \ce{NH2} also react with C-bearing species to form CN,
eventually leading to \ce{N2} formation by Reaction (\ref{react:n+cn}).
Figure \ref{fig:nh3_phd} shows the impact of \ce{NH3} photodesorption on the conversion of \mbox{\ion{N}{i}} into \ce{N2}.
The solid lines represent our fiducial model, while the dashed lines represent the model, 
in which the products of \ce{NH3} photodesorption are set to be N + 3H.
The comparison between the two models indicates that the formation of nitrogen hydrides in the gas phase triggered by \ce{NH3} photodesorption  
accelerates the conversion of \mbox{\ion{N}{i}} into \ce{N2}.
Note that in the context of dark cloud chemistry where interstellar UV radiation field is neglected, 
\ce{NH} and \ce{NH2} are produced from \ce{N2} via ion-neutral gas-phase chemistry \citep[e.g.,][]{hilyblant10}.

\ce{N2} ice is formed via the adsorption of gaseous \ce{N2}.
Reaction between \ce{N2} and atomic hydrogen is significantly endothermic, $>$10$^4$ K \citep[][and references therein]{hidaka11}, 
and thus \ce{N2} ice does not react with icy H atoms.

\subsection{Ammonia deuteration}
\begin{figure}
\resizebox{\hsize}{!}{\includegraphics{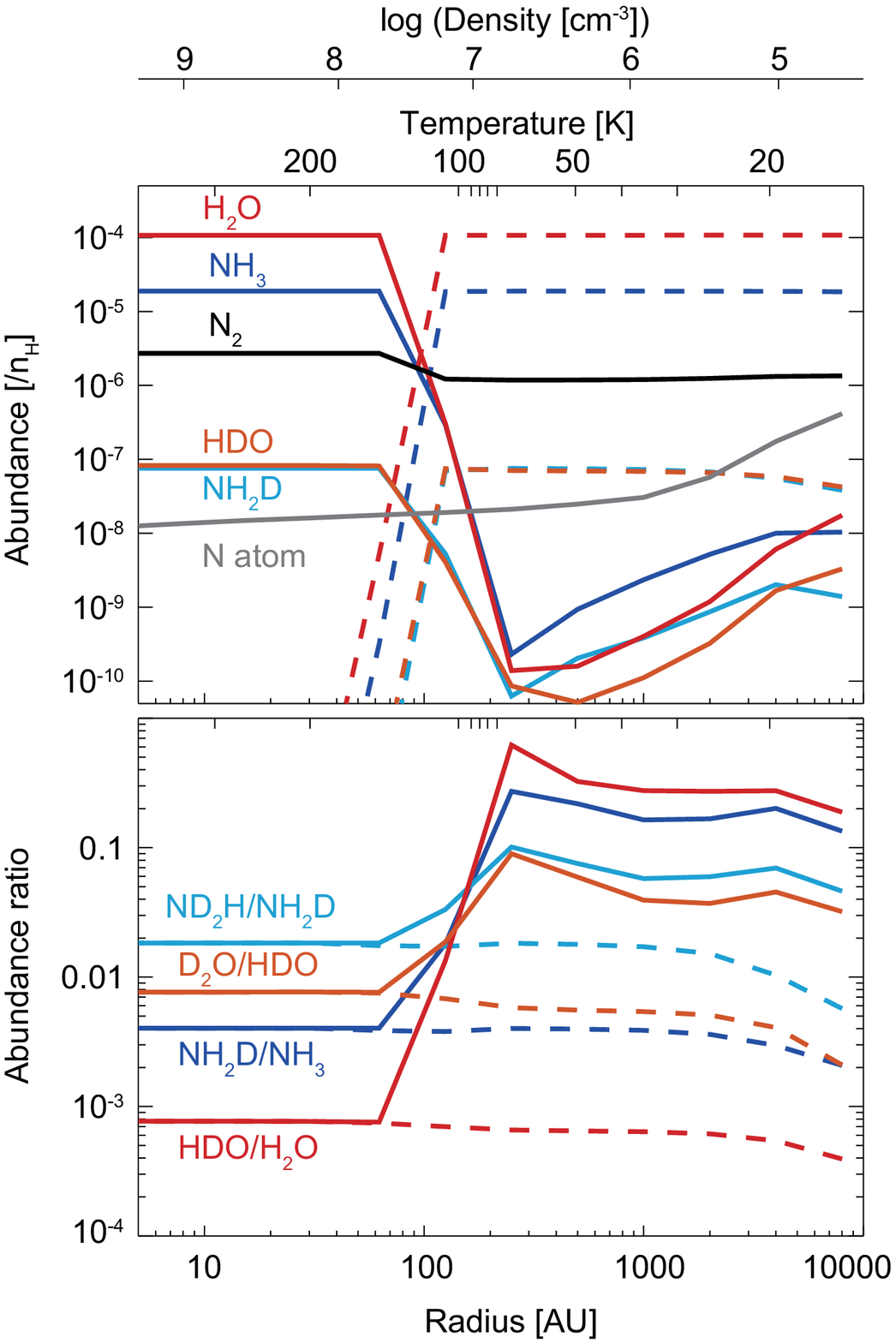}}
\caption{Abundances of selected species with respect to hydrogen nuclei (top) and the abundance ratios (bottom) in the protostellar envelope  
in the fiducial model at $9.3 \times 10^4$ yr after the protostellar birth. The labels at the top represents the temperature and density structures.
The solid lines represent gaseous species, while the dashed lines represent species in the whole ice mantle.
}
\label{fig:a2b_envelope}
\end{figure}

Figure \ref{fig:a2b_envelope} shows the radial profiles of the abundances of selected species with respect to hydrogen nuclei (top panel) and 
the deuterium fractionation ratios in water and ammonia (bottom panel) in the protostellar envelope.
Both icy water and ammonia sublimate into the gas phase at $R \lesssim 100$ AU at temperatures of $\gtrsim$150 K in our model.
\ce{N2} trapped in the ice mantle also sublimates into the gas phase with the sublimation of water and ammonia ices. 

The deuteration ratios of ammonia and water in the gas phase of the protostellar envelope have two common characteristics in our model, 
both of which have been observationally confirmed for water \citep{coutens12,coutens14,persson13}.
First, the abundance ratio between singly-deuterated and non-dueterated forms in the inner warm regions ($T \gtrsim 150$ K) 
is smaller than that in the outer cold envelope (e.g., $4\times10^{-3}$ versus 0.2 for ammonia in our model).
The deuteration ratio in the outer envelope is determined by the gas-phase ion-neutral chemistry, 
while that in the warm regions is dominated by ice sublimation \citep[see also][]{aikawa12,taquet14}.
Second, in the warm regions, the abundance ratio between doubly-deuterated and singly-deuterated forms is higher than 
that between singly-deuterated and non-deuterated forms (e.g., [$\nddh$]/[$\nhhd$] $\sim$ 4).
This is consistent with our prediction from the analytical two-stage model;
in the case when most of atomic nitrogen is consumed during ice formation stages, 
the [$\nddh$]/[$\nhhd$] ratio in the whole ice mantle is greater than unity.
The [$\ddo$]/[$\hdo$] ratio in the inner warm regions is ten in our model, which is consistent with our previous work \citep{furuya16}.
The molecular abundances and the deuterium fractionation ratios in the warm regions are summarized in Table \ref{table:summary}.

\renewcommand\thefootnote{\alph{footnote}} 
\begin{table*}
\caption{Summary of Model Results}
\label{table:core}
\centering
\begin{tabular}{ccccccccccc}
\hline\hline
\multirow{2}{*}{Model}\footnotemark[1] & \multirow{2}{*}{\mbox{\ion{N}{i}} (\%)\footnotemark[2]} &
\multirow{2}{*}{\ce{N2} (\%)\footnotemark[2]} & 
\multirow{2}{*}{\ce{NH3} (\%)\footnotemark[2]} & 
\raisebox{-0.75em}{$\dfrac{\ce{NH3}}{\ce{H2O}}$} & 
\raisebox{-0.75em}{$\dfrac{\ce{NH2D}}{\ce{NH3}}$} & 
\raisebox{-0.75em}{$\dfrac{\ce{ND2H}}{\ce{NH2D}}$} &
\raisebox{-0.75em}{$\dfrac{\ce{ND2H}/\ce{NH2D}}{\ce{NH2D}/\ce{NH3}}$} &
\raisebox{-0.75em}{$\dfrac{\ce{HDO}}{\ce{H2O}}$} &
\raisebox{-0.75em}{$\dfrac{\ce{D2O}/\ce{HDO}}{\ce{HDO}/\ce{H2O}}$} &
\raisebox{-0.75em}{$\dfrac{\ce{CH3OD}}{\ce{CH3OH}}$} \\

\hline
A\footnotemark[3]   &  5.1(-2) & 22 & 76 & 1.8(-1) & 4.0(-3) & 1.8(-2) & 4.6 & 7.7(-4) & 10 & 1.4(-2)\\
B   &  0.1     & 51 & 46 & 1.1(-1) & 1.0(-2) & 1.8(-2)  & 1.8 & 7.6(-4) & 9.5 & 1.3(-2)\\
C   & 57      & 13 & 28 & 6.5(-2) & 2.0(-2) & 1.3(-2) & 6.6(-1) & 7.9(-4) & 8.3 & 1.3(-2)\\
A'\footnotemark[4]  &  6.4(-2) & 29 & 69 & 1.6(-1) & 9.6(-3) & 1.8(-2) & 1.9  & 7.9(-4) & 9.4 & 1.4(-2)\\
\hline \\
\end{tabular}
\tablefootnote{a}{Values in the warm gas ($T>150$ K) at $9.3\times10^4$ yr after the protostellar birth. $a(-b)$ means $a\times10^{-b}$.}
\tablefootnote{b}{Fractions of elemental nitrogen in percentage form.}
\tablefootnote{c}{Our fiducial model.}
\tablefootnote{d}{Similar to model A, but without Reaction (\ref{react:nh+h2}) and the same reactions of the corresponding deuterium isotopologues.}
\label{table:summary}
\end{table*}

\renewcommand{\thefootnote}{\arabic{footnote}}
\setcounter{footnote}{0}

Figure \ref{fig:layer}, panel (c) shows the abundances of non-deuterated and deuterated forms of water ice (black) and ammonia ice (blue) 
normalized by their maximum abundances.
The formation of \ce{NH3} ice is behind that of \ce{H2O} ice (in terms of the analytical two-stage model, $\alpha$ for nitrogen is smaller than that for oxygen), 
while the timing of the formation of deuterated ammonia ice and deuteratd water ice are similar.
These are reflected in the lower [$\nddh$]/[$\nhhd$] ratio than the [$\ddo$]/[$\hdo$] ratio in the warm regions in our fiducial model.
Also the delayed \ce{NH3} ice formation is reflected in the $\nhhd$ ratio in our fiducial model;
the $\nhhd$ ratio is between the $\hdo$ ratio and the \ce{CH3OD}/\ce{CH3OH} ratio (but see Section \ref{sec:consideration}).

\section{Parameter dependences}
\subsection{Primary nitrogen reservoir}
In our fiducial model, most nitrogen is locked in \ce{NH3} ice (76 \% of overall nitrogen and the \ce{NH3}/\ce{H2O} abundance ratio of $\sim$18 \%), 
as widely seen in published gas-ice astrochemical models \citep[e.g.,][]{chang14,pauly16}.
In this subsection, we discuss the effect of the primary nitrogen reservoir on the [$\nddh$]/[$\nhhd$] ratio.

\begin{figure*}
\resizebox{\hsize}{!}{\includegraphics{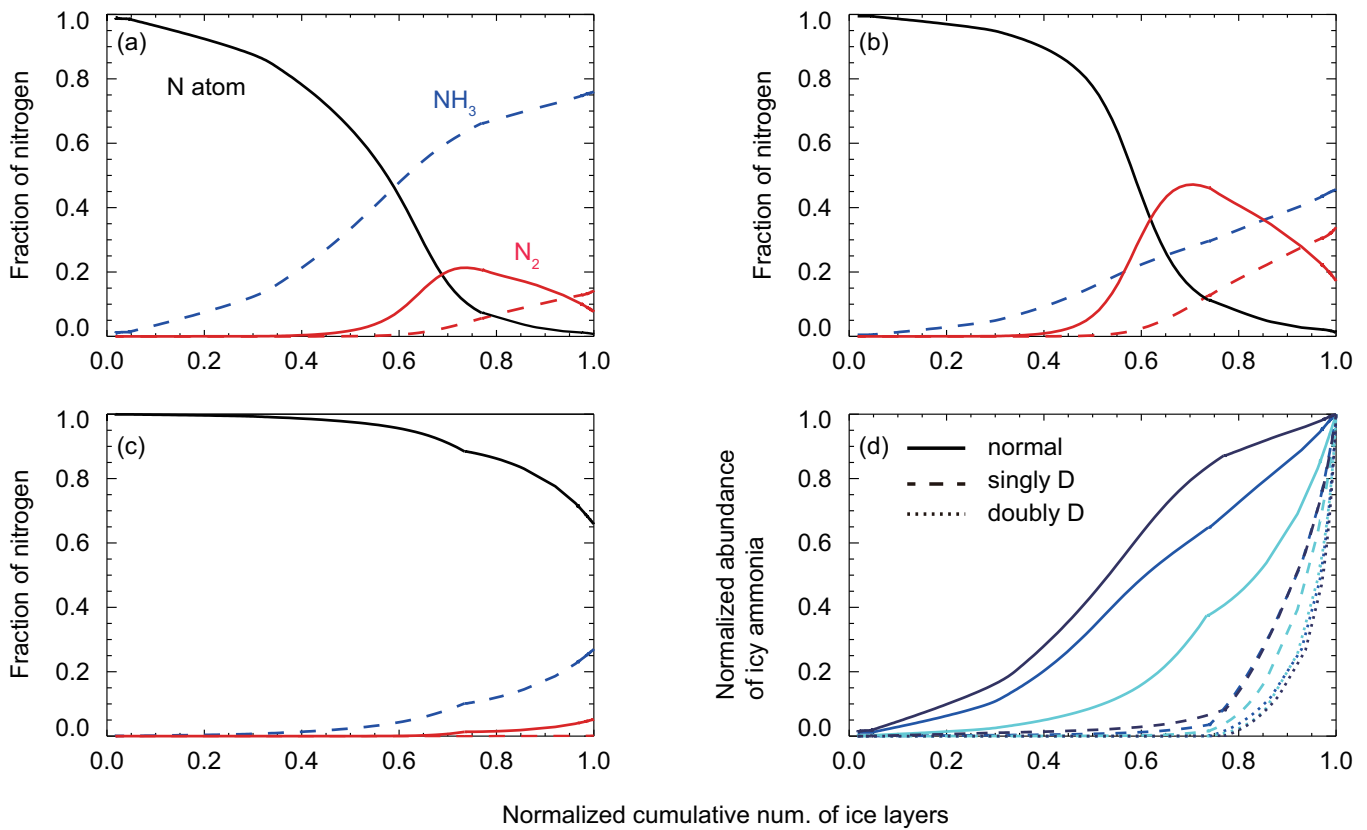}}
\caption{Fraction of elemental nitrogen in \mbox{\ion{N}{i}}, \ce{N2}, and \ce{NH3} in the fiducial model (panel a), model B (panel b), and model C (panel c) 
in the same fluid parcel shown in Figure \ref{fig:layer}
as functions of the cumulative number of ice layers normalized by the maximum number of ice layers.
The solid lines represent gaseous species, while the dashed lines represent icy species.
Panel (d) shows abundances of non-deuterated and deuterated forms of ammonia ice normalized by their maximum abundances in the fiducial model (navy), 
model B (blue), and model C (cyan).
Note that the maximum number of ice layers and the maximum abundances of ammonia ices are different among the models.
}
\label{fig:budget}
\end{figure*}

As discussed in Section \ref{sec:n_partition}, photodesorption of \ce{NH3} assists the conversion of \mbox{\ion{N}{i}} into \ce{N2},
while it slows down the accumulation of \ce{NH3} ice.
Then it is expected that the fraction of nitrogen locked in \ce{NH3} ice decreases with increasing the photodesorption yield of \ce{NH3}, 
while that of \ce{N2} increases.
The recently measured photodesorption yield for pure \ce{NH3} ice in laboratory is $2.1^{+2.1}_{-1.0}\times10^{-3}$ per incident photon \citep{martin17}.
We run the model in which the ptotodesortption yield of \ce{NH3} is $3\times10^{-3}$ (i.e., 3 times larger than in our fiducial model).
Hereafter we refer to this model as model B.
Figure \ref{fig:budget} compares the fractions of elemental nitrogen locked in the selected species in our fiducial model (panel a) and those in model B (panel b). 
The horizontal axis is the cumulative number of layers formed in the ice mantle 
normalized by the total number of ice layers at the end of ice formation (70-80 MLs depending on a model).
As has been expected, the fraction of elemental nitrogen locked in \ce{N2} is higher in model B than in the fiducial model (51 \% versus 22 \%).
In model B, \ce{NH3} ice, gas and icy \ce{N2}, and \mbox{\ion{N}{i}} contain 46 \%, 51 \%, and 0.1 \% of overall nitrogen, respectively.
Partitioning of nitrogen between \ce{NH3} ice and \ce{N2} is sensitive to the photodesorption yield of \ce{NH3} ice.

Both in the fiducial model and in model B, most nitrogen is locked in molecules.
The analytical model presented in Section \ref{sec:theory} predicts that the [$\nddh$]/[$\nhhd$] ratio is similar to the statistical value of 1/3 
and lower than unity when \mbox{\ion{N}{i}} remains as the primary nitrogen reservoir.
In order to simulate this case, we run an additional model (labeled model C).
In model C, the rates of any reactions which include \mbox{\ion{N}{i}} as a reactant are reduced by a factor of 20.
Then in model C, both the formation of \ce{NH3} ice and \ce{N2} gas are slowed down compared to that in the fiducial model.
This assumption is very artificial and model C should be considered as just a numerical experiment. 
In model C, \ce{NH3} ice, gas and icy \ce{N2}, and \mbox{\ion{N}{i}} contain 28 \%, 13 \%, and 57 \% of overall nitrogen, 
respectively (Table \ref{table:summary} and Figure \ref{fig:budget}, panel c).

The [$\nddh$]/[$\nhhd$] ratio is the largest in the fiducial model (4.6) followed in order by model B (1.8) and model C (0.66).
The conversion of \mbox{\ion{N}{i}} to \ce{NH3} ice and \ce{N2} occurs earlier in the fiducial model than in model B, 
and the significant fraction of nitrogen remains as \mbox{\ion{N}{i}} even at the end of the ice formation stage in model C (Figure \ref{fig:budget}, panels a,b,c).
Thus, as predicted by the analytical stage model, the evolution of the \mbox{\ion{N}{i}} abundance is reflected in the [$\nddh$]/[$\nhhd$] ratio.
Figure \ref{fig:budget}, panel d shows when non-deuterated and deuterated forms of ammonia ice are mainly formed 
in the fiducial model, in model B, and in model C.
The fraction of ammonia ice without significant deuterium fractionation ($\lesssim$0.8 in the normalized cumulative number of ice layers) 
is the largest in the fiducial model and the smallest in model C 
(in terms of the analytical two-stage model, $\alpha$ is the smallest in the fiducial model and the largest in model C).

\subsection{Additional considerations}
\label{sec:consideration}
In our fiducial model, the main formation reaction of \ce{NH3} ice is the barrier-mediated reaction \ce{NH2} + \ce{H2} $\rightarrow$ \ce{NH3} + H
on a grain surface rather than the barrierless reaction \ce{NH2} + \ce{H} $\rightarrow$ \ce{NH3}.
\citet{furuya15} found that the level of water ice deuteration depends on which formation reaction, the barrier-mediated reaction 
\ce{OH} + \ce{H2} $\rightarrow$ \ce{H2O} + H or the barrierless reaction \ce{OH} + \ce{H} $\rightarrow$ \ce{H2O} is more effective.
When the barrier-mediated reaction is the primary formation reaction, water ice deuteration can be significantly suppressed  
through the cycle of photodissociation by interstellar UV radiation and reformation of water ice, which efficiently removes deuterium from
water ice chemistry \citep[see also][]{kalvans17}.
We confirmed that the same mechanism is at work for ammonia as well as water.
In our fiducial model, the $\nhhd$ ratio in the icy species ($\sim$10$^{-4}$-10$^{-3}$) is smaller than 
the atomic D/H ratio in the gas phase ($\sim$10$^{-3}$-10$^{-2}$) in the cloud formation stage by a factor of $\gtrsim$10, 
while the difference becomes much smaller in the core stage (Figure \ref{fig:layer}, panels b,d).
While laboratory experiments demonstrated that \ce{OH} + \ce{H2} $\rightarrow$ \ce{H2O} + \ce{H} proceeds on a cold surface by quantum tunneling \citep{oba12},
to the best of our knowledge, \ce{NH2} + \ce{H2} $\rightarrow$ \ce{NH3} + H on a cold surface has not been studied in laboratory.

In order to check the dependence of our results on the main formation reaction of \ce{NH3} ice, 
we rerun our fiducial model without the surface reaction \ce{NH2} + \ce{H2} and the same reactions of the corresponding deuterium isotopologues.
Compared to the fiducial model, the $\nhhd$ ratio in the warm gas ($T \gtrsim 150$ K) is enhanced by a factor of $\sim$2, and close to the \ce{CH3OD}/\ce{CH3OH} ratio.
The [$\nddh$]/[$\nhhd$] ratio is reduced by a factor of $\sim$2, but still much higher than that in model C, in which \mbox{\ion{N}{i}} remains as the primary reservoir of nitrogen.
The main formation reaction of \ce{NH3} ice does not affect our results qualitatively.

\citet{fedoseev15a} performed laboratory experiments for ammonia ice formation in CO ice through sequential hydrogenation of N atoms.
They found that HNCO forms in their experiments likely via the reaction,
\begin{align}
\ce{NH} + \ce{CO} \rightarrow \ce{HNCO}, \label{react:nhco}
\end{align}
where NH is the intermediate in the formation of \ce{NH3}.
Similarly formamide (\ce{NH2CHO}) may be formed via the following surface reaction \citep{fedoseev16},
\begin{align}
\ce{NH2} + \ce{H2CO} \rightarrow \ce{NH2CHO} + \ce{H}. \label{react:nh2cho}
\end{align}
These two reactions, which are not included in our surface chemical network, may reduce the production rate of ammonia ice on CO-rich layers 
and thus could enhance the [$\nddh$]/[$\nhhd$] ratio.

Recent observations of the warm gas around low-mass protostar IRAS 16293-2422B \citep{coutens16} found that
the levels of deuteration in HNCO and \ce{NH2CHO} are similar to the \ce{CH3OD}/\ce{CH3OH} ratio.
This finding is consistent with the scenario that HNCO and \ce{NH2CHO} are formed via surface reactions on CO-rich ice layers \citep{coutens16}.
The observed \ce{HNCO} and \ce{NH2CHO} column densities are, however, lower than the \ce{CH3OH} column density 
in the same source by factors of $\sim$1000 \citep{coutens16,jorgensen16}.
According to infrared ice observations, the \ce{NH3}/\ce{CH3OH} ratio in the cold outer envelope of low-mass protostars is the order of unity \citep{oberg11}. 
These indicate that either atomic nitrogen is already poor when the catastrophic CO freeze out happens or 
only the small fraction of atomic nitrogen adsorbed on grain surfaces contributes to the formation of HNCO and \ce{NH2CHO} ices.
In either case, Reactions (\ref{react:nhco})-(\ref{react:nh2cho}) would not affect our results qualitatively.

\section{Discussion: Observability of deuterated ammonia in the warm gas}
\label{sec:discussion}
There have been several observational studies to quantify the ammonia deuteration in star formation regions. Most measurements are of the cold gas in prestellar cores and in outer envelopes of deeply embedded protostars, using single dish telescopes \citep[e.g.,][]{lis02,roueff05,harju17}.
For example, observations toward the prestellar core I16293E shows ratios of $\nhhd \approx \nddh \approx 20$ \% \citep{loinard01,roueff05}. Measurements for the cold large scale envelope toward a deeply embedded low-mass protostar NGC 1333-IRAS 4A show a $\nhhd$ ratio of $\sim$7 \% \citep{shah01}.
These extremely high $\nhhd$ and $\nddh$ ratios are reasonably well reproduced in the outer cold regions of our protostellar envelope model ($\gtrsim$1000 AU, Figure \ref{fig:a2b_envelope}).

While in agreement with the modeling presented here, these previous measurements, however, would not reflect the deuteration in ice, being supported by astrochemical models \citep[Figure \ref{fig:a2b_envelope}, see also][]{aikawa12,taquet14}. The nitrogen deuterium fractionation in the cold gas in the large scale envelope do not differ significantly between the various models. The ratio deduced in the large scale envelope is thus not a suitable indicator for distinguishing between the different scenarios of atomic nitrogen evolution. To measure the ammonia deuteration in the bulk ice, interferometric observations towards the recently sublimated warm gas in the inner regions of protostars are necessary. It has been already shown that the values of water deuteration measured in the large scale cold envelope is much higher than that measured in the warm envelope around protostars, where water ice has sublimated \citep{coutens12,persson13}. 

To assess the possibility of distinguishing between the different models with observations, we here calculate the spectra of \ce{NH3}, \ce{NH2D} and \ce{ND2H} toward a typical Class~0 protostar assuming the ratios in model~A in Table~\ref{table:summary}. The synthetic spectra are compared to typical sensitivities in the relevant Karl G Jansky Very Large Array (VLA) and Atacama Large Millimeter/submillimeter Array (ALMA) bands with reasonable integration times (10~hours with VLA and 4~hours with ALMA) calculated using officially available tools 
(sensitivity calculators\footnote{VLA \url{https://obs.vla.nrao.edu/ect/}, ALMA \url{https://almascience.eso.org/proposing/sensitivity-calculator}}) at a target resolution of 0\farcs4. This can then aid the planning of future observations.

To this date, very few detections of compact ammonia emission toward a deeply-embedded protostar exists. \citet{choi07,choi10} presented detections of the $2_{2,0a}-2_{2,0s}$ and $3_{3,0a} - 3_{-3,0s}$ transitions of \ce{NH3} using the VLA telescope at high resolution ($\sim0\farcs3$) toward the protobinary source IRAS~4A, located at a distance of 235~pc \citep{hirota08}. In Fig.~\ref{fig:a2b_envelope} the relevant volume hydrogen density in the warm region is typically $>10^{8}$~cm$^{-3}$, which is well above the critical density for all \ce{NH3} lines \citep[e.g.,][]{maret09}, which shows that LTE is prevalent in these sources and scales. We can then use general relations e.g., \citet{goldsmith99} to reproduce the observed line fluxes. A similar method to constrain an abundance is presented in e.g.\ \citet{persson13,coutens16}. The detected ammonia spectral lines are well reproduced by a column density of $6-7\times10^{17}$~cm$^{-2}$ at an excitation temperature of 250--300~K, assuming LTE and accounting for the optical depth. For Class~I sources, \citet{sewilo17} presented archival observations of ammonia toward the source HH111/HH121 at medium spatial resolution ($\sim8$\arcsec). The sparse results on compact ammonia emission highlights even further the importance of more observations constraining the ammonia abundances in these warm inner regions of deeply embedded protostars.

With observations of \ce{H2^18O} \citep{persson13,persson14} we can estimate the \ce{NH3}/\ce{H2O} column density ratio in the warm gas\footnote{Assuming a \ce{^16O}/\ce{^18O} ratio of 560 \citep[relevant for these galactocentric distances][]{wilson94,wilson99}}, that can be compared with that obtained by infrared ice observations. For the north-western source of the IRAS~4A binary the \ce{H2O} column density was constrained by \citet{persson14} to $4.4\times10^{18}$~cm$^{-2}$, and we get a \ce{NH3}/\ce{H2O} column density ratio in the warm gas of 13-16~\%. This value is close the \ce{NH3}/\ce{H2O} ice column density ratio in low-mass protostellar envelopes, 3-10~\%, depending on sources \citep[][and references therein]{boogert15}, indicating that our LTE method is reasonable. Note that IRAS~4A is a highly extincted Class 0 source and ice measurements are not available.

To estimate the line fluxes of \ce{NH3}, \ce{NH2D} and \ce{ND2H} we will use the deeply embedded source IRAS~16293-2422 (hereafter I16293). It is located in the $\rho$ Ophiucus star forming complex, at a distance of 120~pc \citep{loinard08} and is a binary source with a 5\arcsec\ (600~AU) separation where the north western source is referred to as A, and the south easter as B. The molecular content of I16293 has been extremely well explored with both interferometers and single dish telescopes \citep{bottinelli04, bisschop08, jorgensen11}. \citet{jorgensen16} presents an extensive review of the work done towards I16293. The water isotopologue \ce{H2^18O} was observed at high resolution with ALMA and SMA \citep{persson13}, and it was only possible to deduce the column density toward source A. Using the derived water column density and the \ce{NX3}/\ce{H2O} abundance ratio in our models (X is H or D; Table \ref{table:summary}), we can estimate the \ce{NX3} column density in I16293. Assuming the conservative scenario of model~A from Table~\ref{table:summary}, we here estimate the column densities of \ce{NH3}, \ce{NH2D}, and \ce{ND2H}. The \ce{H2O} water column density from \citet{persson13} is $5.3\times10^{20}$~cm$^{-2}$. Assuming an excitation temperature of 200~K towards the warm gas of I16293 as in the innermost regions of the model (c.f.\ Fig.~\ref{fig:a2b_envelope}), similar to e.g.\ \citet{persson13} we can calculate the expected line strengths.
The water column density gives a column density of $9.5\times10^{19}$~cm$^{-2}$ for \ce{NH3}, $3.8\times10^{17}$~cm$^{-2}$ for \ce{NH2D}, and $6.9\times10^{15}$~cm$^{-2}$ for \ce{ND2H}. 
For the modeled spectra, we assume a size of the emitting region of 0\farcs4 and a beam size of 0\farcs4.

The \ce{NH3}v=0 transition frequencies and constants are taken from the Spectral Line Atlas of Interstellar Molecules (SLAIM), available at \url{http://www.splatalogue.net} \citep[F.~J.~Lovas,][]{remijan07} queried through the Astroquery interface \citep{ginsburg17}. For \ce{NH2D} and \ce{ND2H} the frequencies and constants are from the Cologne Database for Molecular Spectroscopy \citep[CDMS,][]{muller05}.

\subsection{Lines in the 1 to 52~GHz frequency range}
The VLA receivers operate mainly in 8 different bands with a combined continuous spectral coverage between 1 and 50~GHz\footnote{This excludes band 4 and the P-band, operating below 1~GHz, but with only on a limited number of antennae and at a highest angular resolution of $\sim$5\farcs6, which is not relevant for this study.}. With a schedule of different configurations it achieves resolutions ranging from 0\farcs043 to 46\arcsec. ALMA band 1 covers the frequency range 35 to 52~GHz (50-52~GHz is on best-effort basis). The typical 3$\sigma$ sensitivity of an observation run of about 10~hours for VLA is 1~mJy in a 0\farcs4 beam, valid between around 4 and 48~GHz\footnote{Assuming 27 antennae, winter conditions, 2.2~km~s$^{-1}$ channel width, and an elevation of the target between 50--90~degrees.}. For ALMA band 1 it is not possible to estimate the sensitivity with the online estimator, but it is assumed to be at a similar level to VLA ($\sim$1~mJy/beam), but with less observation time.
\begin{figure}
	\resizebox{\hsize}{!}{\includegraphics{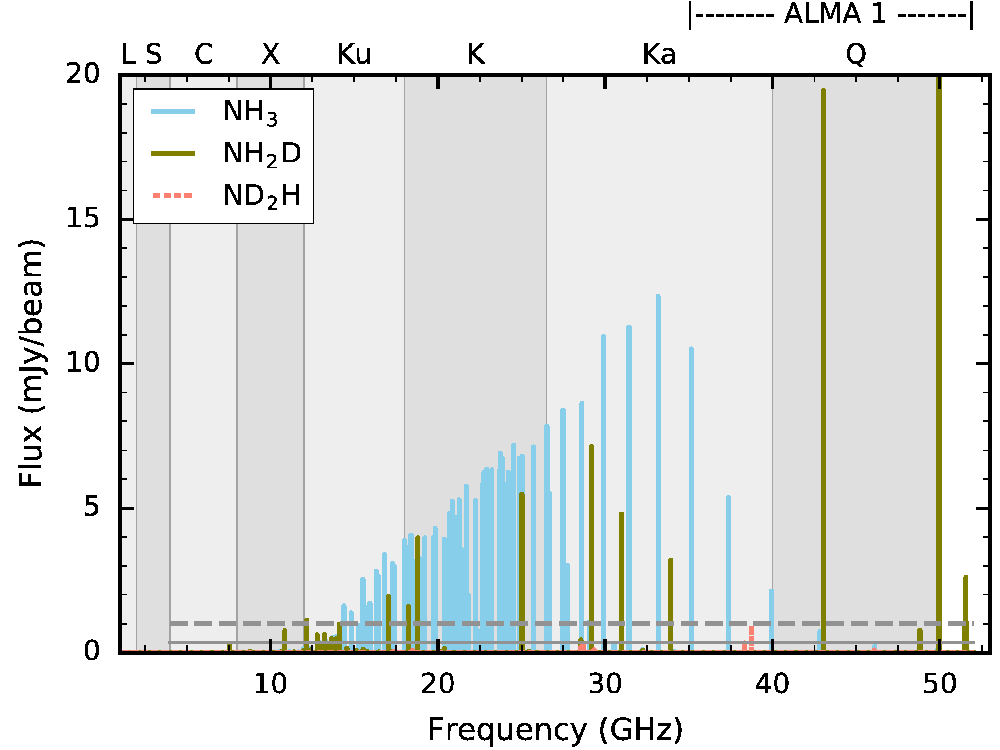}}
	\caption{Synthetic spectrum of \ce{NH3}, \ce{NH2D}, and \ce{ND2H} in the VLA frequency range. The model assumes $T_\mathrm{ex}=200$~K, a source size of 0\farcs4, and a beam of 0\farcs4. Letters at top indicate the relevant VLA band, and where "ALMA 1" indicates the region of ALMA band 1 (35-52~GHz). The solid and dashed line shows the 1 and 3$\sigma$ levels respectively.
	}
	\label{fig:vlaspectrum}
\end{figure}
The resulting LTE spectra for \ce{NH3}, \ce{NH2D} and \ce{ND2H} in the VLA spectral range (1--50~GHz) is shown in Figure~\ref{fig:vlaspectrum}. The main lines to target with VLA are between 15--38~GHz for \ce{NH3}, and while 7 lines of \ce{NH2D} are available in the same range, there are no \ce{ND2H} lines predicted to be above the $3\sigma$ level. The \ce{NH2D} line at 49.96~GHz is just at the edge of the Q-band where sensitivity is significantly lower. Furthermore, as optical depth might be an issue for some of the transitions, a wide spread in $E_\mathrm{u}$ and Einstein~A coefficient is the best strategy. The beam is matched to the rough size of the warm inner regions of the sources, once abundances and emission extent has been determined more exact with high sensitivity observations, it is possible to observe the lines at higher spatial and spectral resolution to constrain the distribution and kinematic origin of ammonia further. As can be seen only a few lines of deuterated ammonia are available with VLA, and they are spread out over several bands which is not an ideal observing strategy. Furthermore, no \ce{ND2H} transitions can be detected with VLA, thus we have to turn to higher frequencies and observe the target with the ALMA telescope. ALMA band 1 covers a few \ce{NH3} and \ce{NH2D} lines, and could thus be used for observations, once operational. However, the main transition bands are covered by VLA and is thus the most suitable instrument.

\subsection{Lines 65 to 720~GHz frequency range}
The ALMA telescope array is equipped with receivers covering frequencies between 35 and 720 GHz in bands 1 through 9. Band 1 and 2, i.e.\ 35 to 90~GHz is under construction, thus the band 2 sensitivity is assumed to be similar as in band 3 (and band 1 sensitivity similar to VLA at same frequency). With baselines ranging from 160~m to 16~km the resulting spatial resolution range from 0\farcs020 to 4\farcs8. The typical 3$\sigma$ sensitivity of an observation run of about 4~hours integration time in a 0\farcs4 beam differs between the bands, but lies around 0.7~mJy~beam$^{-1}$ in bands 1 through 7, and increases to around 4~mJy~beam$^{-1}$ in band 8 and 5~mJy~beam$^{-1}$ in band 9. However, the high frequency bands (band 8 and 9) need lower levels of precipitable water vapor (PWV) for these sensitivities. Thus they are less favorable since less time is available for such observations, and there are transitions at lower frequencies.

As the figures show, there are several lines of \ce{NH2D} and \ce{ND2H} present well above the sensitivity level in most receiver bands. However the highest overlapping density of lines with high signal-to-noise (SNR) is in band~7. Thus the best target for deuterated ammonia with ALMA is in band~7, around 300~GHz. The higher frequency bands, band 8 and 9 need better weather for a given sensitivity, and is as such less efficient to observe. Furthermore, given the high SNR of the lines in band~7, it is possible to observe sources with lower column densities.

While the main frequency region for \ce{NH3} is in the VLA receiver bands, there are two lines in ALMA band 7 that might be possible to detect (at 354 and 361~GHz). However, their high energy levels ($E_\mathrm{u}\approx800$~K) and low Einstein A ($log_{10}(A_{ij})\approx-8$) makes them more difficult to detect. In addition to the target lines of \ce{NH2D} and \ce{ND2H}, there are two \ce{ND3} lines as well; the \ce{ND3} $1_{0,0}-0_{0,1}$ transition at 307~GHz, and the $1_{0,1}-0_{0,0}$ transition at 310~GHz. Thus with a single spectral setup and around 4 hours of ALMA integration time in band 7, it should be possible to characterize the \ce{NH2D}, \ce{ND2H} and \ce{ND3} column densities, all at once. This further highlights the importance of band 7 for ALMA observations of ammonia. Combining this with VLA observations of \ce{NH3} it is possible to constrain the ratios studied here, and constrain the amount of \mbox{\ion{N}{i}} in young, deeply embedded protostars. Once again, targeting a combination of transitions with different energy levels and transition probabilities is the best strategy to constrain the abundances.

\begin{figure}
	\resizebox{\hsize}{!}{\includegraphics{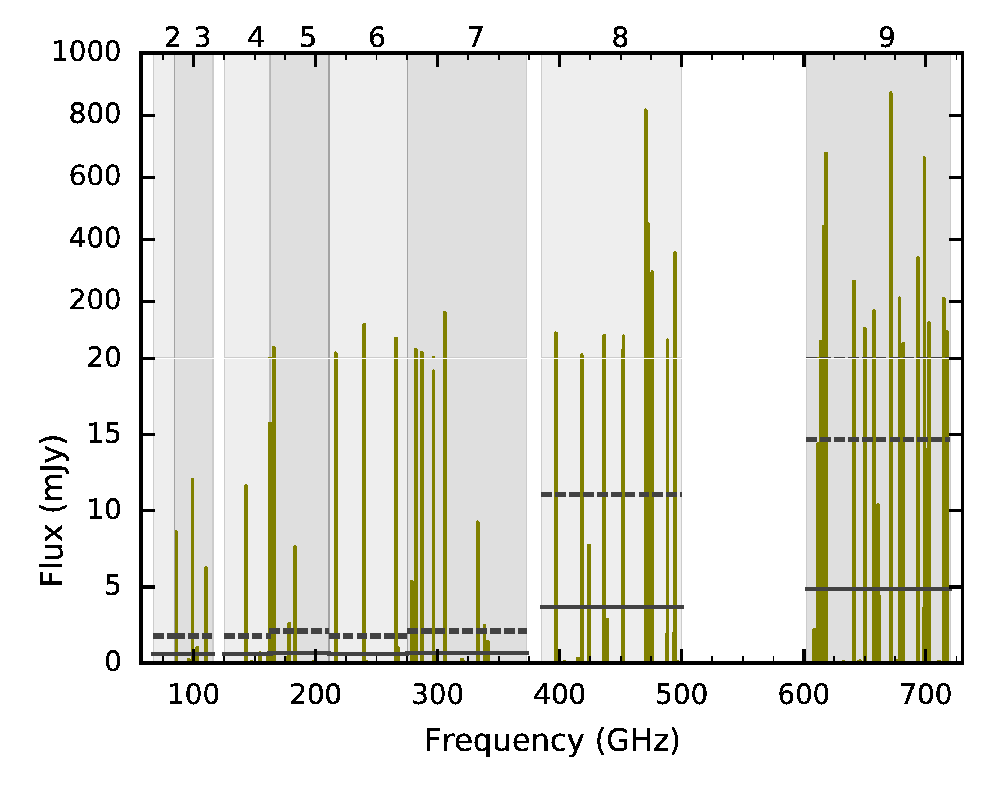}}
	\caption{Synthetic spectrum of \ce{NH2D} in the ALMA frequency range. The model assumes $T_\mathrm{ex}=200$~K, a source size of 0\farcs4, and a beam of 0\farcs4. The numbers at the top indicate the relevant ALMA band. Note the change in scale of the flux around 20~mJy. The solid and dashed lines shows the 1 and 3$\sigma$ levels respectively.
	}
	\label{fig:alma_nh2d_spectrum}
	\resizebox{\hsize}{!}{\includegraphics{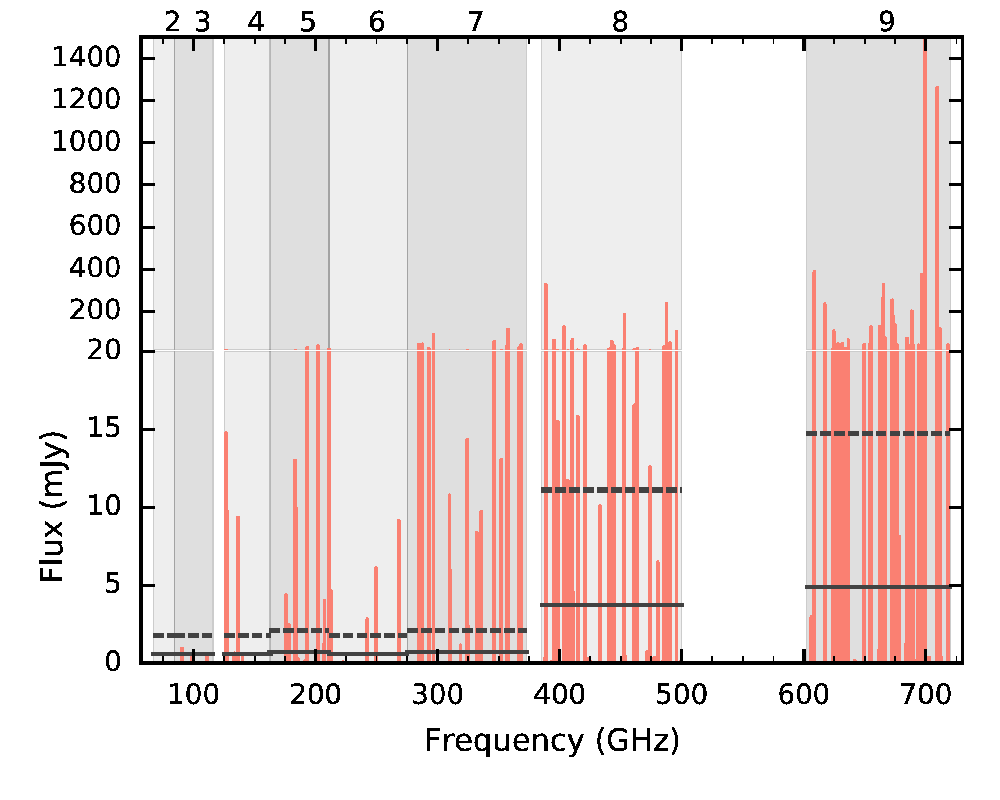}}
	\caption{Synthetic spectrum \ce{ND2H} in the ALMA frequency range. The model assumes $T_\mathrm{ex}=200$~K, a source size of 0\farcs4, and a beam of 0\farcs4. The numbers at the top indicate the relevant ALMA band. Note the change in scale of the flux around 20~mJy. The solid and dashed lines shows the 1 and 3$\sigma$ levels respectively.
	}
	\label{fig:alma_nd2h_spectrum}
\end{figure}

\section{Conclusion}
\label{sec:summary}
Partitioning of elemental nitrogen in star-forming regions is not well constrained.
Most nitrogen is expected to be partitioned among \mbox{\ion{N}{i}}, \ce{N2}, and icy N-bearing molecules, such as \ce{NH3} and \ce{N2}.
Neither \mbox{\ion{N}{i}} nor \ce{N2} is directly observable in the cold gas, while the \ce{N2} abundance can be 
constrained via a proxy molecule, \ce{N2H+}.
In this paper, we have proposed an indirect way to constrain the amount of atomic nitrogen in the cold gas of star-forming clouds, 
via deuteration in ammonia ice.
Using a simple analytical model and gas-ice astrochemical simulations, which trace the evolution from the formation of molecular clouds to protostellar cores, 
we showed the evolution of the \mbox{\ion{N}{i}} abundance in the cold ($\sim10$ K) ice formation stages is reflected in the icy [$\nddh$]/[$\nhhd$] ratio.
If \mbox{\ion{N}{i}} remains as the primary reservoir of nitrogen during cold ice formation stages, 
the ratio is close to the statistical value of 1/3 and lower than unity, 
whereas if \mbox{\ion{N}{i}} is largely converted into N-bearing molecules, the ratio should be larger than unity.
The [$\nddh$]/[$\nhhd$] ratio in ice mantles in star-forming clouds can be quantified with VLA and ALMA observations of 
the inner warm regions around protostars, where ammonia ice has completely sublimated, with reasonable integration times (Section \ref{sec:discussion}).

We also found that partitioning of nitrogen between \ce{NH3} ice and \ce{N2} is sensitive to the photodesorption yield of \ce{NH3} ice.
The fraction of elemental nitrogen locked in \ce{NH3} ice decreases with increasing the photodesorption yield of  \ce{NH3} ice,
while that in \ce{N2} increases.
The increased efficiency of \ce{NH3} photodesorption slows down the accumulation of \ce{NH3} ice, 
while it assists the conversion of \mbox{\ion{N}{i}} into \ce{N2}; 
photodesorbed \ce{NH3} is further photodissociated in the gas phase and the photofragment, \ce{NH2} or \ce{NH}, reacts 
with \mbox{\ion{N}{i}} to form \ce{N2} (Section \ref{sec:n_partition}).

Finally, as demonstrated in this paper for nitrogen and in \citet{furuya16} for oxygen, 
multiple deuteration can be a strong tool to constrain the evolution of atomic reservoirs in the cold gas.
The method could be applied to other elements, such as sulfur, by observing e.g., H$_2$S, H$_2$CS and their deuterated forms, 
if they are mainly formed by surface reactions.
However, to do that, laboratory and theoretical studies of their formation and deuteration pathways on surfaces are necessary.

\section*{Acknowledgements}
We thank Satoshi Yamamoto and Yuri Aikawa for fruitful discussions on \ce{N2} chemistry.
We also thank the referee for the valuable comments that helped to improve the manuscript.
K.F. acknowledges the support by JSPS KAKENHI Grant Number 17K14245. M.V.P. postdoctoral position is funded by the ERC consolidator grant 614264.









\bsp	
\label{lastpage}
\end{document}